\documentclass[journal]{IEEEtran}
\usepackage{cite}

\ifCLASSINFOpdf
\else
\usepackage[dvips]{graphicx}
\fi
\usepackage{amsmath}

\usepackage{flushend}
\usepackage{color}
\usepackage{graphicx}

\begin{document}
%
\title{Dielectric insulated transmission lines in receiving antenna operation}
%
%
%

\author{Reuven Ianconescu,~\IEEEmembership{}
        Vladimir Vulfin ~\IEEEmembership{}
\thanks{Reuven Ianconescu is with the Department
of Electrical Engineering, Shenkar college of engineering and design, Ramat Gan,
Israel, e-mail: riancon@gmail.com}
\thanks{Vladimir Vulfin is with Electromagnetics Infinity LTD, Israel, email: vladimir.vulfin@eminfinity.com}
}

\maketitle

\begin{abstract}
This work derives exact expressions for the voltage induced into a two conductors dielectrically isolated
transmission line by a monochromatic incident plane wave from an arbitrary direction, at a given polarization. The transmission line
cross section, consisting of the conductors and the dielectric material, may be of any shape, provided the cross section size is much smaller than
the wavelength, so that the waves in radiation mode satisfy the quasi TEM condition. We calculate analytically the voltage along the transmission
line for given end loads and compare the results with ANSYS HFSS simulation results. Our calculations are based on the knowledge of the radiation
from such a transmission line, derived elsewhere and the radiation-absorption reciprocity.

\end{abstract}


%
\IEEEpeerreviewmaketitle

\section{Introduction}
%
%
%
%

\IEEEPARstart{I}{n} two previous works we made full analyses of the radiation
properties of transmission lines (TLs), including radiated field intensity and polarization,
radiation pattern, radiation resistance and more. In \cite{full_model_arxiv} we analyzed TL in free space and
in \cite{QuasiTEM} TL isolated by dielectrics. The results of \cite{full_model_arxiv} have been used to extrapolate the inverse problem, i.e.
the response of free space TL to a monochromatic incident plane wave from an arbitrary direction, at a given polarization. The results of this work
have been published in \cite{receive_TL}.

\begin{figure}[!tbh]
\includegraphics[width=9cm]{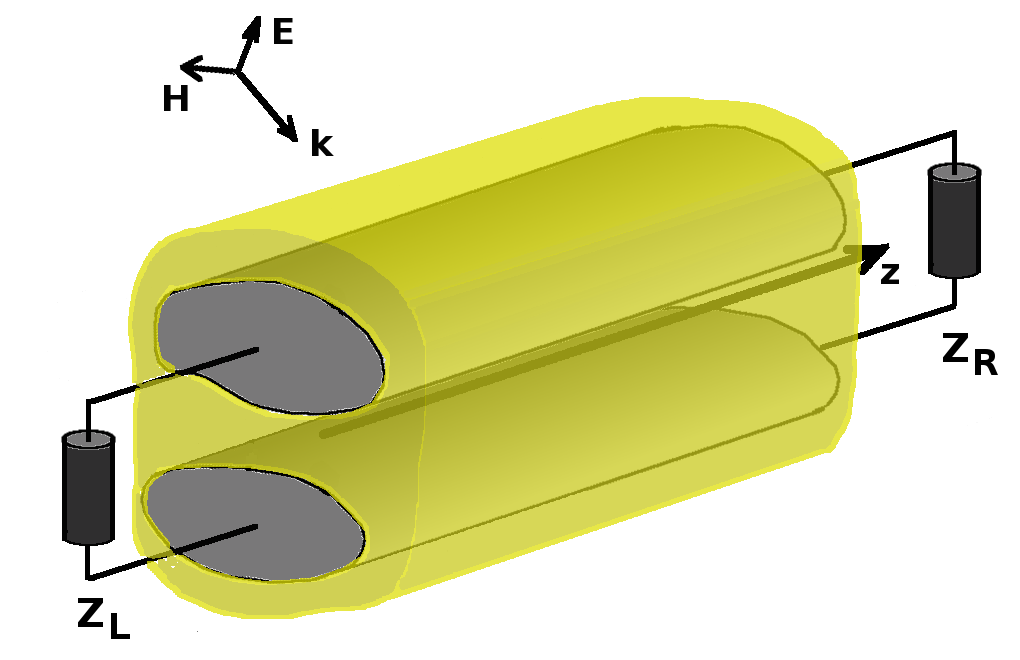}
\caption{Configuration of a two ideal conductors transmission line (TL), connected at both sides
to passive loads: $Z_L$ (left) and $Z_R$ (right), hit by a monochromatic plane wave propagating
toward the center of coordinates.
The cross section is electrically small and may be of any shape.
The loads are located at the terminations of the TL, and are shown farther away, due to technical drawing
limitations. We consider the most general case, so that the plane wave hits from any arbitrary direction.}
\label{config}
\end{figure}

In the current work we use the results of \cite{QuasiTEM} to investigate the inverse problem of a dielectric isolated TL connected to passive
loads absorbing electromagnetic power, namely we calculate the voltage (amplitude and phase) developed on a two-conductors dielectrically isolated TL hit by a
monochromatic plane wave, as shown in Figure~\ref{config}.
Similarly to \cite{QuasiTEM}, we consider ideal lossless TL of any small electric cross section.

\begin{figure}[!tbh]
\begin{centering}
\includegraphics[width=5cm]{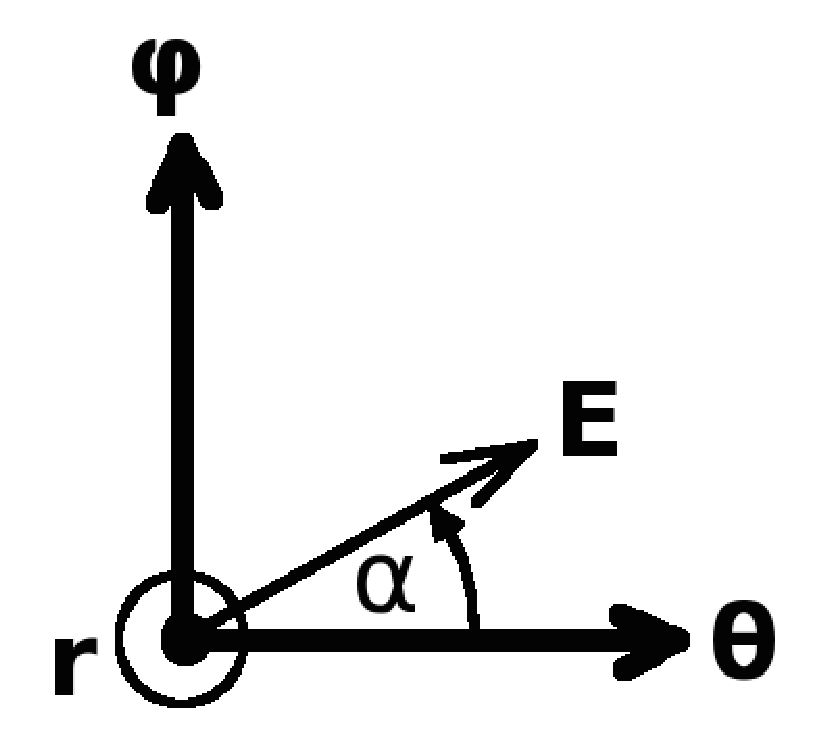}
\caption{The incident plane wave propagates toward the center of coordinates in the $-\mathbf{\widehat{r}}$
direction. In 
spherical coordinates the local equiphase surface is $\theta,\varphi$ and the
polarization is at angle $\alpha$ from the $\theta$ axis, so that at the origin
the E field is given by Eq.~(\ref{E_plane_wave}). This is the most general case of linear polarization.}
\label{plane_wave}
\end{centering}
\end{figure}

A more detailed look at the incident plane wave and its polarization is given in Figure~\ref{plane_wave}.
It propagates toward the coordinates origin with phase
$e^{-j\mathbf{k}\cdot\mathbf{r}}$, so that the wavenumber vector $\mathbf{k}=-\mathbf{\widehat{r}}k$ points toward
the origin. Expanding $\mathbf{\widehat{r}}$
in Cartesian unit vectors, the phase can be written as
\begin{equation}
e^{jk[x\sin\theta\cos\varphi+y\sin\theta\sin\varphi+z\cos\theta]},
\label{plane_wave_phase}
\end{equation}
where $\theta$ and $\varphi$ are the spherical angles which represent the direction of the plane wave incidence.
The plane wave travels toward the center of coordinates, perpendicular to the $\theta,\varphi$ plane,
with a polarization angle $\alpha$ from the $\theta$ axis, so that the E field at the origin, where its phase
is defined to be 0 (Eq.~(\ref{plane_wave_phase})) is given by
\begin{equation}
\mathbf{E}=E_0(\boldsymbol{\widehat{\theta}}\cos\alpha+\boldsymbol{\widehat{\varphi}}\sin\alpha)
\label{E_plane_wave}
\end{equation}
or its components
\begin{equation}
E_{\theta}=E_0\cos\alpha \,\,\,\,\,\, ; \,\,\,\,\,\, E_{\varphi}=E_0\sin\alpha.
\label{E_plane_wave_components}
\end{equation}

The configuration as defined here is a scattering problem, requiring a full wave solution to set the tangential component of the electric field to 0 on the surface of the perfect conductors and boundary conditions on the dielectric. However, we derive here an analytic solution to this problem, using the S parameters methodology we used in \cite{receive_TL} for free space TL. Like in \cite{receive_TL} the analytic results are compared with a full wave HFSS (high-frequency structure simulator) solution.

The work is organized as follows. In Section~\ref{radiation_properties} we bring a conclusive summary of the radiation properties of quasi TEM dielectrically isolated TL, based on the findings in \cite{QuasiTEM}. We present the far E-field for several cases and the main parameters which affect the radiation.

In Section~\ref{Voltage_S_matrix} we present the derivation of the main results of this work: the voltage along a quasi TEM dielectrically isolated TL hit by a monochromatic plane wave. We define ports along the TL and two far antenna ports, and use the results shown in Section~\ref{radiation_properties} to derive a generalized scattering matrix. Using this matrix we derive in subsection~\ref{DM} the voltage on a TL matched at both ends (double matched) and after that in \ref{arbitrary} we generalize the result for a TL terminated by any loads.

In Section~\ref{HFSS} we describe the full wave HFSS simulations performed, and explain some delicate issues regarding
the measurements of the voltage. We describe the simulated cross section and the parameters obtained from it.

In Section~\ref{validation_hfss} we validate the analytic results with full wave HFSS simulations for the cross section described above and compare the theoretical results obtained in Sections~\ref{Voltage_S_matrix} with the full wave HFSS solutions. In subsection~\ref{matched} we compare results for double matched TL and in subsection ~\ref{non_matched} we compare several non matched cases.

The work is ended with some concluding remarks.

\section{Radiation properties of quasi TEM transmission lines}
\label{radiation_properties}

To fulfill the purpose of this work we need some knowledge about the radiation from dielectric isolated TL developed in \cite{QuasiTEM}. A TL extending from $-L$ to $L$ in the $z$ direction carrying the current $I^+e^{-j\beta z}$ (defined to the right in the ``upper'' conductor) radiates a far E-field given in spherical coordinates $\theta$ and $\varphi$ by
\begin{align}
  \mathbf{E}^+=&-2\eta_0 kG(r)I^+d \sin[kL(n_{eq}-\cos\theta)] \notag \\
             &[\boldsymbol{\widehat{\varphi}}A\sin\varphi+\boldsymbol{\widehat{\theta}}B\cos\varphi].
\label{E_plus_dielectric}
\end{align}
Here $\eta_0=377 [\Omega]$ is the free space impedance, $k=2\pi f/c$ is the free space wavenumber, $G(r)=\frac{e^{-jkr}}{4\pi r}$ is the free space Green function and $n_{eq}=\sqrt{\epsilon_{eq}}=\beta/k$ is the equivalent refraction index, which defines the relation between the TL wavenumber $\beta$ and the free space wavenumber $k$. As shown in Appendix~1 of \cite{QuasiTEM}, those also define the relation between the capacity per unit length with and without dielectric: $C=\epsilon_{eq} C_{\text{free space}}$, and the characteristic impedance with and without dielectric: $Z_0=Z_{0\,\text{free space}}/n_{eq}$.

The parameter $d$ represents the vector distance (or dipole length) between the ‘negative’ and ‘positive’ conductor in a twin lead equivalent representation. We always choose the $x$ axis in the dipole direction, therefore we treat $d$ as scalar. The method to obtain $d$ for a free space TL via a cross section analysis has been described in Appendix~B of \cite{full_model_arxiv} and has been generalized for a dielectric isolated TL in Appendix~2 of \cite{QuasiTEM}. Finally the parameters $A$ and $B$ are
\begin{equation}
A=\frac{\overline{n}-\cos(\theta)}{n_{eq}-\cos(\theta)} \,\,\,\,\,\, B=\frac{1-\overline{n}\cos(\theta)}{n_{eq}-\cos(\theta)}.
\label{A_B}
\end{equation}
We encounter here a new parameter $\overline{n}\equiv n_{eq}/\epsilon_{p}$, where $\epsilon_{p}$ is a relative permittivity related to the geometry of the transverse polarization currents. The physical meaning of $\epsilon_{p}$, and its connection to $\epsilon_{eq}$ is given in Appendix~3 of \cite{QuasiTEM}, and we bring here a short summary of this issue.

The polarization current density $J_p$ is related to the displacement current density $J_d$ by $\frac{J_p}{J_d}=\frac{\epsilon_r-1}{\epsilon_r}$ in the dielectric and 0 in the air, so that the average relation over all the E-field lines is $\left\langle\frac{J_p}{J_d}\right\rangle=\frac{\epsilon_{eq}-1}{\epsilon_{eq}}$, and this can be regarded as a definition of $\epsilon_{eq}$. As shown in Appendix~2 of \cite{QuasiTEM}, the polarization current density component in the dipole direction (i.e. the $x$ direction) constitute a radiating current element proportional to $\left\langle\frac{J_{p\,x}}{J_d}\right\rangle$, which is smaller than $\left\langle\frac{J_p}{J_d}\right\rangle$. The relative permittivity $\epsilon_p$ is defined by this equality: $\left\langle\frac{J_{p\,x}}{J_d}\right\rangle=\frac{\epsilon_p-1}{\epsilon_p}$.

For a geometry of parallel E-field (similar to parallel plates), $J_d=J_{d\,x}$, hence $\epsilon_{p}=\epsilon_{eq}$, so in general $\epsilon_{p}\le\epsilon_{eq}$. For more twin lead like geometries, a significant part of the E-field is in the $\pm y$ direction, and those cancel out their contributions, so that we have a smaller component in the $x$ direction. However because the E-field terminates at the conductors, their projection on $x$ cannot be 0 everywhere. If it could be, we would have $\epsilon_{p}=1$, i.e. no contribution from polarization currents, so 1 is an unattainable infimum of $\epsilon_p$, and not a minimum. Hence the limits of $\epsilon_p$ and $\overline{n}= n_{eq}/\epsilon_{p}$ are
\begin{equation}
1<\epsilon_{p}\le\epsilon_{eq}=n_{eq}^2\,\,\,\,\,\,;\,\,\,\,\,\, 1/n_{eq} \le \overline{n} < n_{eq}
\label{eps_p}
\end{equation}
so that $\overline{n}=n_{eq}$ is an unattainable supremum of $\overline{n}$. This supremum cannot be tested with ANSYS HFSS simulation on a real geometry, but it has been compared in \cite{QuasiTEM} with a theoretical work that ignored the polarization currents \cite{Nakamura_2006}, giving identical results.

Similarly for a TL extending from $-L$ to $L$ in the $z$ direction carrying the current $I^-e^{j\beta z}$ (defined to the right in the ``upper'' conductor) radiates a far E-field given in spherical coordinates $\theta$ and $\varphi$ by
\begin{align}
  \mathbf{E}^-=&-2\eta_0 kG(r)I^-d \sin[kL(\cos\theta+n_{eq})] \notag \\
             &[-\boldsymbol{\widehat{\varphi}}C\sin\varphi+\boldsymbol{\widehat{\theta}}D\cos\varphi]
\label{E_minus_dielectric}
\end{align}
where the parameters $C$ and $D$ are
\begin{equation}
C=\frac{\overline{n}+\cos(\theta)}{n_{eq}+\cos(\theta)} \,\,\,\,\,\, D=\frac{1+\overline{n}\cos(\theta)}{n_{eq}+\cos(\theta)}
\label{C_D}
\end{equation}

\section{Derivation of the voltage along the TL}
\label{Voltage_S_matrix}



\subsection{The voltage along a double-matched TL}
\label{DM}
\begin{figure}[!tbh]
\includegraphics[width=9cm]{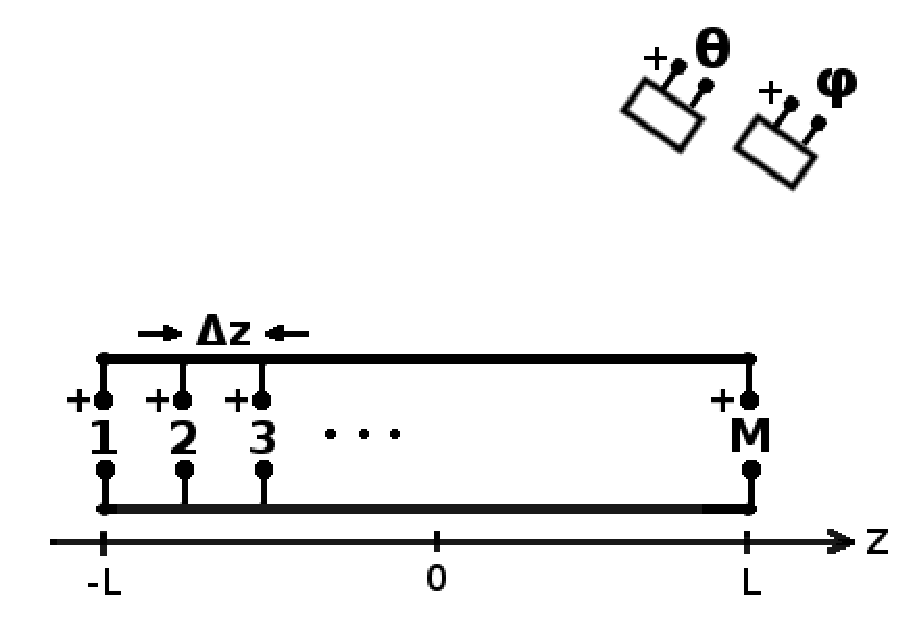}
\caption{Circuit with $M$ parallel ports along the TL at distances $\Delta z$.
Ports 1 and $M$ are defined for the TL impedance $Z_0$, and the middle ports 2 .. $M-1$ are defined for a high reference impedance $Z_H$.
Two additional ports representing far antennas matched for the $\boldsymbol{\widehat{\theta}}$ and
$\boldsymbol{\widehat{\varphi}}$ polarizations are defined for the reference impedance identical to the TL characteristic impedance $Z_0$. The two antenna ports are named $\theta$ and $\varphi$.}
\label{voltage_S}
\end{figure}

An analytic solution for the voltage along a TL hit by a monochromatic plane wave is possible with the aid of the circuit shown in Figure~\ref{voltage_S}, and our aim is to derive first the voltage on a TL matched on both sides (double-matched). The circuit has parallel ports along the TL (numbered 1 to $M$), and two additional ports which represent far antennas (named $\theta$ and $\varphi$). We will derive a generalized scattering matrix for this circuit, with different reference port impedances: the TL ports 1 and $M$ as well as the antenna's ports are defined for the reference impedance $Z_0$ (i.e. the characteristic impedance of the TL), while the middle ports 2 to $M-1$ are defined for a very high reference impedance $Z_H\to\infty$, so that a matched port is open circuit. Due to this choice, when feeding the circuit with incoming voltages $V^+_{\theta}$ and/or $V^+_{\varphi}$ the outgoing voltages $V^-_{1,..M}$ represent the actual voltage at each port location:
\begin{equation}
V_n=V^-_n=S_{n,\theta}V^+_{\theta}+S_{n,\varphi}V^+_{\varphi}\equiv V_{\textrm{DM}}
\label{matched_voltage_on_TL}
\end{equation}
and given $\Delta z$ is arbitrarily small (or $M$ arbitrarily large) this result is the voltage as function of $z$
along the double-matched TL for the given incident plane wave. The ``DM'' subscript stands for double-matched.

In case we excite this circuit with $V^+_1$ (all other ports matched) we have a forward wave in the TL, so that the far E field due to this wave is given in Eq.~\ref{E_plus_dielectric}, while in case we excite the circuit with $V^+_M$ (all other ports matched) we have a backward wave in the TL, resulting in a far E-field given in Eq.~\ref{E_minus_dielectric}. If we excite a middle port (2 to $M$) and all other ports matched, we have a combination of the above far fields, so in all cases we know the far field hitting the antennas and to calculate the scattering matrix elements we need to {\it translate} the E-field into voltages $V^-_{\theta}$ and $V^-_{\varphi}$.

This ``translation'' has to be consistent with the inverse case in which we excite the circuit with $V^+_{\theta}$ and/or $V^+_{\varphi}$ (all other ports matched). Those antennas being far, excite a plane wave of intensity $E_0$ in the vicinity of the TL, as described in the introduction, shown in Figure~\ref{config} and detailed in Figure~\ref{plane_wave}.

As evident from (\ref{E_plus_dielectric}) and (\ref{E_minus_dielectric}) the radiated far fields are proportional to the dipole length
$d$ (or equivalent separation distance in a twin lead representation), explained in the previous section. Hence the voltage response of the TL in receive mode is also proportional to this separation distance $d$, we therefore define the incoming voltage associated with the plane wave:
\begin{equation}
V^+\equiv E_0\,d,
\label{Voltage_E_field_equivalence}
\end{equation}
or by components (equivalent to Eq.~(\ref{E_plane_wave_components})):
\begin{equation}
V^+_{\theta}=V^+\cos\alpha \,\,\,\,\,\, ; \,\,\,\,\,\, V^+_{\varphi}=V^+\sin\alpha.
\label{Voltage_wave_components}
\end{equation}
The translation of the radiated E-field into voltages $V^-_{\theta}$ or $V^-_{ \varphi}$, consistent with Eq.~\ref{Voltage_E_field_equivalence} has been derived in Appendix~A of \cite{QuasiTEM}, and is given by
\begin{equation}
V^-_{\theta \,\text{or}\, \varphi}=\frac{Z_0 E_{\theta\, \text{or}\, \varphi}}{2jk\eta_0G(r)d},
\label{scale_factor}
\end{equation}
where $Z_0$ in the numerator is due to defining the $\theta$ and $\varphi$ ports in Figure~\ref{voltage_S} for the reference impedance identical to the characteristic impedance of the TL.

As evident from Eq.~\ref{matched_voltage_on_TL} to obtain the voltage on the double-matched TL we need only the $(n,\theta)$ and $(n,\varphi)$ elements of $S$ (for $1\le n\le M$)). We will therefore calculate the $(\theta,n)$ and $(\varphi,n)$ elements and use reciprocity.

\begin{figure}[!tbh]
\includegraphics[width=9cm]{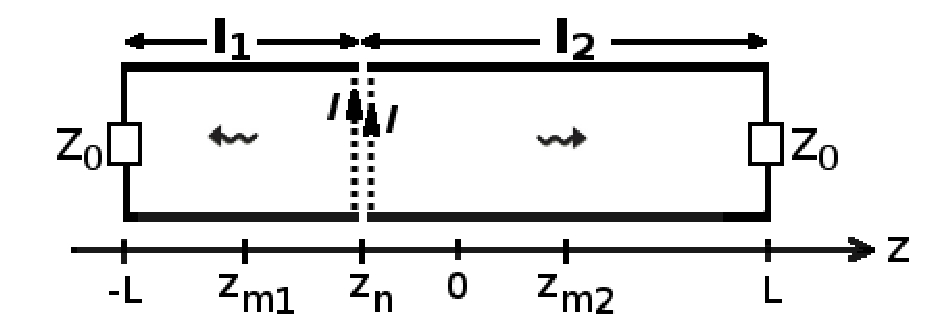}
\caption{The circuit described in Figure~\ref{voltage_S} while fed at the middle port $n$ by $V^+_n$ and matched at
all other ports. Defining $I\equiv V^+_n/Z_H$, we get two ``separate'' TLs, the one between $[z_n,L]$ carrying
a forward wave $Ie^{-j\beta(z-z_n)}$ and the one between $[-L,z_n]$ carrying a backward wave $-Ie^{j\beta(z-z_n)}$.}
\label{middle_feed_voltage}
\end{figure}

 For this, we shall use the $\theta$ and $\varphi$ components
of the far E field radiated from the circuit in Figure~\ref{voltage_S} fed at port $n$,
all other ports matched. This configuration is shown in Figure~\ref{middle_feed_voltage}.
We define the lengths from port $n$ to the terminations:
\begin{equation}
l_1=z_n+L\,\,\,\,\,\, ; \,\,\,\,\,\,l_2=L-z_n,
\label{l_12_k}
\end{equation}
and the middle points $z_{m_1}$, $z_{m_2}$:
\begin{equation}
z_{m_1}=-l_2/2\,\,\,\,\,\, ; \,\,\,\,\,\,z_{m_2}=l_1/2,
\label{z_m12}
\end{equation}
Given all other ports (except $n$) are matched, the circuit develops a forward wave {\bf only} in the region $[z_n,L]$
and a backward wave only in the region $[-L,z_n]$, shown schematically in the figure. Hence the feeding source
at port $n$ ``sees'' an impedance $Z_0$ from each side of the TL.
To calculate the far E field radiated by the circuit in Figure~\ref{middle_feed_voltage} we
use the results in \cite{QuasiTEM}, summarized in Eqs.~(\ref{E_plus_dielectric}) and (\ref{E_minus_dielectric})
for the forward and backward waves, respectively.
We set into Eqs.~(\ref{E_plus_dielectric}) $I^+\to \frac{V^+_n}{Z_H}e^{-j\beta l_2/2}e^{jkz_{m_2}\cos\theta}$ and $2L\to l_2$,
and into Eqs.~(\ref{E_minus_dielectric}) $I^-\to -\frac{V^+_n}{Z_H}e^{-j\beta l_1/2}e^{jkz_{m_1}\cos\theta}$ and $2L\to l_1$.
Adding the results, yields the far E field:
\begin{equation}
\eta_0G(r)2kd \frac{V^+_n}{Z_H}[-\boldsymbol{\widehat{\varphi}}\sin\varphi (Cf_1+Af_2)+ \boldsymbol{\widehat{\theta}}\cos\varphi(Df_1-Bf_2)],
\label{E_from_V_conf}
\end{equation}
where the functions $f_1$ and $f_2$ are defined as:
\begin{align}
&f_1\equiv e^{-jk(n_{eq}l_1+l_2\cos\theta)/2}\sin[kl_1(n_{eq}+\cos\theta)/2] \notag \\
&f_2\equiv e^{-jk(n_{eq}l_2-l_1\cos\theta)/2}\sin[kl_2(n_{eq}-\cos\theta)/2],
\label{f_12}
\end{align}
and we note that the port $n$ to which the field (\ref{E_from_V_conf}) is related is given indirectly
by the values of $l_{1,2}$ via the functions $f_{1,2}$ by setting $z_n=z$ in (\ref{l_12_k}).

Separating the E field into $\theta$ and $\varphi$ components and scaling with (\ref{scale_factor})
we obtain the outgoing voltage waves
\begin{equation}
V^-_{\theta}=-jV^+_n(Z_0/Z_H)[D f_1-B f_2]\cos\varphi
\label{V_minus_theta}
\end{equation}
\begin{equation}
V^-_{\varphi}=jV^+_n(Z_0/Z_H)[C f_1+ A f_2]\sin\varphi.
\label{V_minus_phi}
\end{equation}
Results (\ref{V_minus_theta}) and (\ref{V_minus_phi}) define the $(\theta,n)$ and
$(\varphi,n)$ elements for $S$, respectively. For the columns $1<n<M$:
\begin{equation}
S_{\theta,\,1<n<M}=j(Z_0/Z_H)[B f_2- D f_1]\cos\varphi
\label{S_theta_k}
\end{equation}
\begin{equation}
S_{\varphi,\,1<n<M}=j(Z_0/Z_H)[C f_1+ A f_2]\sin\varphi.
\label{S_phi_k}
\end{equation}
For column $n=1$ or $M$, the results are similar, only replace $\frac{Z_0}{Z_H}$ by 1.
The transpose elements are found by the reciprocity condition $S_{i,j}Z_j=S_{j,i}Z_i$ (see Appendix~B in \cite{receive_TL} or \cite{pozar}), where $Z_i$ and $Z_j$ are the reference impedances for which ports $i$ and $j$ have been defined, respectively. Given the ports 1 and $M$ are defined for $Z_0$ of the TL and ports 2 .. $M-1$ are defined for $Z_H$, the transposed relations $S_{n,\theta}$ and $S_{n,\varphi}$ are given by Eqs.~(\ref{S_theta_k})
and (\ref{S_phi_k}), under the replacement $\frac{Z_0}{Z_H}\to 1$:
\begin{equation}
S_{n,\theta}=j[B f_2- D f_1]\cos\varphi
\label{S_k_theta}
\end{equation}
\begin{equation}
S_{n,\varphi}=j[C f_1+ A f_2]\sin\varphi.
\label{S_k_phi}
\end{equation}
Now using Eqs.~(\ref{Voltage_wave_components}) and (\ref{matched_voltage_on_TL}), we obtain the voltage on the double-matched TL:
\begin{equation}
V_{\textrm{DM}}=jV^+[(B f_2-D f_1)\cos\varphi\cos\alpha + (A f_2+C f_1)\sin\varphi\sin\alpha]
\label{V_tot}
\end{equation}
This is the final result for the voltage on a TL matched at both sides (double-matched).
\subsection{The voltage along a TL terminated by arbitrary loads}
\label{arbitrary}

We generalize this result for arbitrary impedances at ports 1 and $M$: $Z_L$ (left) at port 1 and $Z_R$ (right) at port $M$, as shown in Figure~\ref{config}. Here the middle ports are still matched (with $Z_H\to\infty$, i.e. open) so we need two additional incoming voltages from ports 1 and $M$ in the summation (\ref{matched_voltage_on_TL}):
\begin{equation}
V^-_n= V_{\textrm{DM}}+S_{n,1}V^+_1+S_{n,M}V^+_M
\label{non_matched_voltage_on_TL}
\end{equation}
We therefore need additional elements of the matrix $S$, specifically the elements $(i,1)$
and $(i,M)$, where $1\le i\le M$. For this calculation port $i=1$ or $M$ is fed by the incoming voltage $V^+_i$, ``seeing'' a total
impedance $Z_0$, so that $V^-_i=0$, i.e. both ports are matched. On the TL a forward/backward wave only arises for $i=1,M$ respectively. Hence the outgoing voltage at port $n$ is
$V^+_ie^{-j\beta|z_n-z_i|}$, specifically: $V^+_ie^{-j\beta l_1}$ for $n=1$ and $V^+_ie^{-j\beta l_2}$ for $n=M$, using $\beta=n_{eq}k$, results in
\begin{equation}
S_{1\le n\le M\,,\,i=1,M}=
\left\{
\begin{array}{l l}
e^{-jn_{eq}kl_1} &  \,\,  i=1\text{ and }n\neq i \\
e^{-jkn_{eq}l_2} &  \,\,  i=M\text{ and }n\neq i \\
0         & \,\,   n=i
\end{array}
\right.
\label{partial_column_i_eq_1_or_M}
\end{equation}
where the value of $n$ is resides in the values of $l_{1,2}$. Now we need to express $V^+_1$ and $V^+_M$ to set them in Eq.~(\ref{non_matched_voltage_on_TL}), hence we define the reflection coefficients $\Gamma_L$ (at the left side) and $\Gamma_R$ (at the right side):
\begin{equation}
\Gamma_L=\frac{Z_L-Z_0}{Z_L+Z_0}\,\,\,\text{and}\,\,\,\Gamma_R=\frac{Z_R-Z_0}{Z_R+Z_0}.
\label{reflection_coeffs}
\end{equation}
so that:
\begin{equation}
V^+_1=\Gamma_L V^-_1\,\,\,\text{and}\,\,\,V^+_M=\Gamma_R V^-_M,
\label{V_plus_1_M}
\end{equation}
therefore Eq.~(\ref{non_matched_voltage_on_TL}) becomes
\begin{align}
V^-_n= V_{\textrm{DM}}+S_{n,1}\Gamma_L V^-_1+S_{n,M}\Gamma_R V^-_M
\label{V_minus_k}
\end{align}
To find $V^-_1$ and $V^-_M$ we set $n=1$ and $n=M$ in Eq.~(\ref{V_minus_k}), obtaining:
\begin{equation}
V^-_1=\Gamma_R V^-_M e^{-jn_{eq}k2L} + V_{\textrm{DM}}(-L),
\label{V_minus_1_1}
\end{equation}
\begin{equation}
V^-_M=\Gamma_L V^-_1 e^{-jn_{eq}k2L} + V_{\textrm{DM}}(L).
\label{V_minus_M_1}
\end{equation}
The solution of the above two equations for $V^-_1$ and $V^-_M$ yields:
\begin{equation}
V^-_1=\frac{\Gamma_R e^{-j2n_{eq}kL}V_{\textrm{DM}}(L)+V_{\textrm{DM}}(-L)}{1-\Gamma_L\Gamma_R e^{-j4n_{eq}kL}} 
\label{V_minus_1_solution}
\end{equation}
\begin{equation}
V^-_M=\frac{\Gamma_L e^{-j2n_{eq}kL}V_{\textrm{DM}}(-L)+V_{\textrm{DM}}(L)}{1-\Gamma_L\Gamma_R e^{-j4n_{eq}kL}}.
\label{V_minus_M_solution}
\end{equation}
Those results can be set now in Eq.~\ref{V_minus_k}, but here we have to be careful. Eq.~\ref{V_minus_k} gives the total voltage only for the middle ports $1<n<M$, for which $V^+_n=0$, while the total voltage on port 1 or $M$ is $V^-_{1,M}+V^+_{1,M}$.

We want to define a correction term $\Delta V$ to add to $V_{\textrm{DM}}$ for the total (non matched) voltage. The last two terms in Eq.~\ref{V_minus_k} constitute this correction term for $1<n<M$. This is written using Eq.~\ref{partial_column_i_eq_1_or_M} for $n\neq i$
\begin{equation}
\Delta V(z)= e^{-jn_{eq}kl_1}\Gamma_L V^-_1 + e^{-jn_{eq}kl_2} \Gamma_R V^-_M.
\label{Delta_V}
\end{equation}

Now using $n=1$ in Eq.~\ref{V_minus_k}, given $S_{1,1}=0$ we have $V^-_1= V_{\textrm{DM}}(-L)+S_{1,M}\Gamma_R V^-_M$. But the total voltage at port 1 is $V^-_1+V^+_1= V_{\textrm{DM}}(-L)+\Gamma_L V^-_1+S_{1,M}\Gamma_R V^-_M$, which results in the same correction given by Eq.~\ref{Delta_V} for $z=-L$, i.e. $l_1=0$. Similarly it is easy to show that Eq.~\ref{Delta_V} holds for port $M$, therefore it holds for all $z$.

To conclude this section, we derived the voltage on a double matched TL in Eq.~\ref{V_tot} and the correction term (\ref{Delta_V}) so that the total voltage for the most general case is
\begin{equation}
V(z)=  V_{\textrm{DM}}(z) + \Delta V(z),
\label{V_NM}
\end{equation}

\section{Full wave HFSS simulations}
\label{HFSS}

We describe in this Section the HFSS simulations done for the scattering problem defined in
Figure~\ref{config}. The results of this simulations are compared in the next section with the
analytic results derived in Section~\ref{Voltage_S_matrix}.

\begin{figure}[!tbh]
\includegraphics[width=8cm]{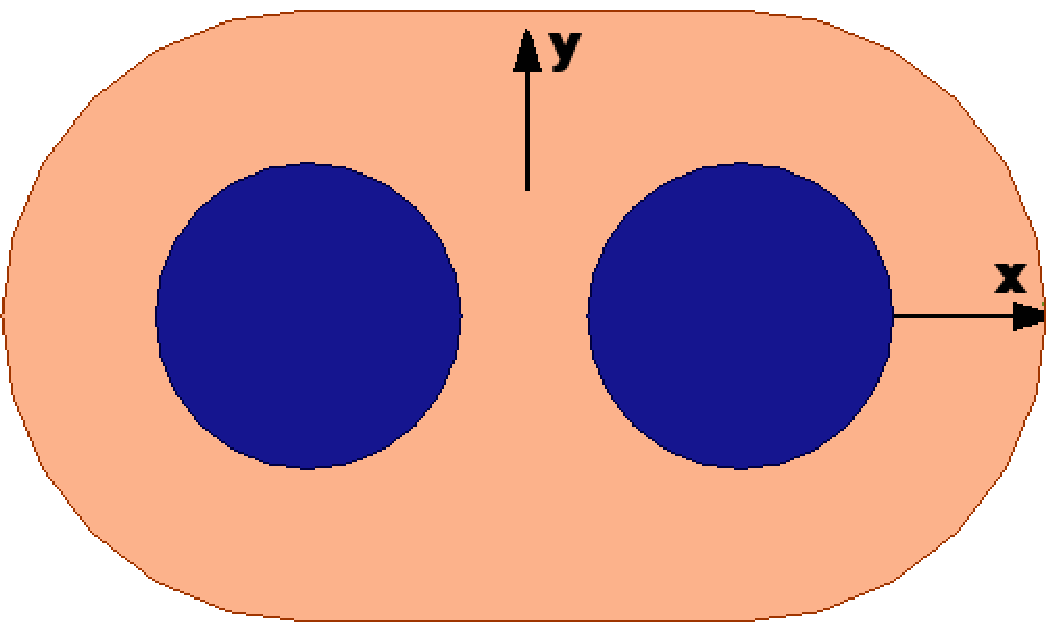}
\caption{The cross section consists of two circular shaped ideal conductors of radius $a=1.27$~cm (dark blue), the distance
between their centres being $s=3.59$~cm. The dielectric insulator (pink) is circular with radius $2a$ for $|x|>s/2$
and rectangular in the region $|x|<s/2$. The relative permittivity of the dielectric insulator is $\epsilon_r=3$.}
\label{cross_section_ins}
\end{figure}

%
We simulated a TL with the cross section shown in Figure~\ref{cross_section_ins}.

The cross section analysis, as described in Appendix~2 of \cite{QuasiTEM}, yields the following parameters for this TL. 
\begin{equation}
d=2.46\text{cm}\,\,\,n_{eq}=1.613\,\,\,\overline{n}=0.85\,\,\,Z_0=65.5\Omega
\label{params}
\end{equation}
To be used in Eq.~(\ref{V_tot}) and Eq.~(\ref{V_NM}) for the comparison with the simulation results.


The electric field of a plane wave is by default $E_0=1$V/m in the HFSS simulation, so to normalize the
results for $V^+=E_0\,d=1$V (see Eq.~\ref{Voltage_E_field_equivalence}) we divide the measured results
by the value of $d$ in Eq.~(\ref{params}).

For convenience, we shall use a fixed TL length of $l\equiv 2L=125$cm, and test for different
frequencies. We measure the voltage along the TL from $z=-61.25$cm to $z=61.25$cm
at intervals of 6.125cm, in total at 21 points. At the TL terminations $z=-L$ and $z=L$, we
use inactive lumped ports defined for the impedance we need at those terminations.

Two-conductors dielectrically isolated TL excited {\it only} at terminations, develop the quasi TEM mode, so that both $E_z$ and $H_z$ are small relative to the transverse fields (see Appendix~2 of \cite{QuasiTEM}), practically allowing a voltage measurement $\int \mathbf{E}\cdot \mathbf{dl}$ on any path in the cross section.

In the case analyzed here, the TL is excited by an external plane wave, therefore, depending on the
incidence of this wave $E_z$ and/or $H_z$ are not necessarly small, we therefore need more careful
definitions for the voltage measurements.

\begin{figure}[!tbh]
\includegraphics[width=8cm]{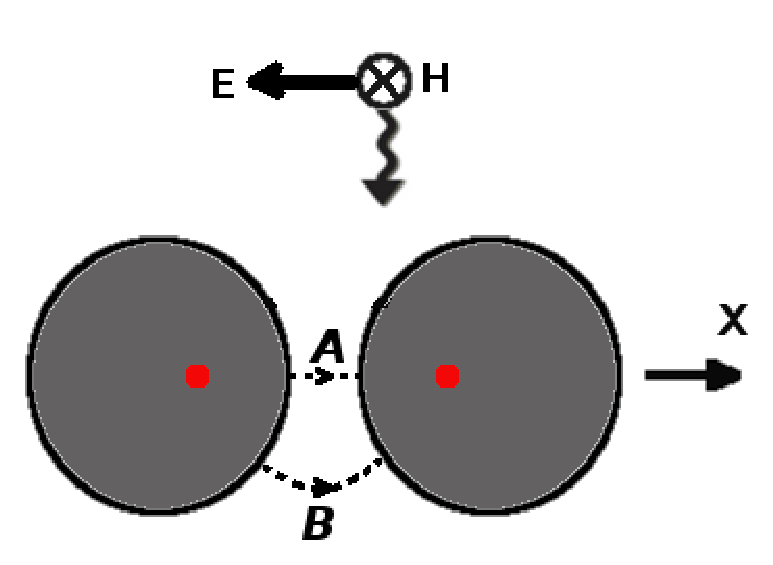}
\caption{Cross section voltage measurement on two possible pathes $A$ and $B$. In case $H_z=0$,
all pathes lead to the same result, namely $\int_A \mathbf{E}\cdot \mathbf{dl}=\int_B \mathbf{E}\cdot \mathbf{dl}$.
However, if $H_z\neq 0$, as for the incident plane wave shown here, the result of the path integral depends on
the paths used, and the correct voltage measurement is $\int_A \mathbf{E}\cdot \mathbf{dl}$, i.e. along the $x$ axis
consistent with the parallel ports definition in Figure~\ref{voltage_S}.}
\label{parallel_cylinders_Hz}
\end{figure}

The voltage measured by the integral $\int \mathbf{E}\cdot \mathbf{dl}$ in the cross
section depends on the chosen integration path if $H_z\neq 0$, as shown in Figure~\ref{parallel_cylinders_Hz}.
To define the correct path we look at the definitions of the parallel ports in Figure~\ref{voltage_S}.
Those have been defined on the $x-z$ plane, so that {\it only} $x$ directed currents flow through the port, and this fact
has been used in the calculation of the S matrix in Section~\ref{Voltage_S_matrix}, see
Figure~\ref{middle_feed_voltage}.

Therefore, to be consistent with the parallel ports definition,
the correct path to measure the voltage is path $A$ (on the $x$ axis) shown in Figure~\ref{parallel_cylinders_Hz},
and this path is used in all the voltage measurements shown in the next section.

\section{Validation of the analytic results}
\label{validation_hfss}

\subsection{Double-matched transmission line}
\label{matched}

We validate in this section the analytic results for a double-matched TL in Eqs~(\ref{V_tot}) by comparison with full wave solution of ANSYS HFSS simulation results, described in the previous section.

We analyse three examples, each from a main incidence direction, by plane waves traveling along the $x$, $y$ or $z$
axes. As mentioned in the previous section, the incident plane wave is scaled to result in $V^+=1$[V] in Eq.~\ref{Voltage_E_field_equivalence}.

\begin{figure}[!tbh]
\includegraphics[width=9cm]{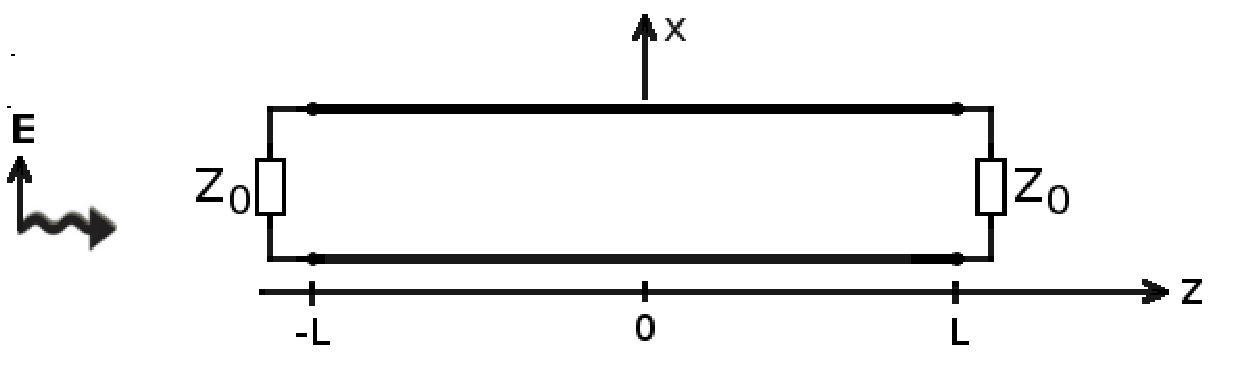}
\caption{Matched TL illuminated by a $\mathbf{\widehat{x}}$ polarised plane wave from $\theta=\pi$.}
\label{left_incidence}
\end{figure}

In the first example we examine a plane wave traveling from $\theta=\pi$, along the $z$ axis, colinear with the TL, having the phase $e^{-jkz}$, as shown in Figure~\ref{left_incidence}. It is easy to check that a $\mathbf{\widehat{y}}$ polarized field yields zero voltage on the TL, we therefore use a $\mathbf{\widehat{x}}$ polarized field.
At $\theta=\pi$, the polarization is writtern as $\mathbf{\widehat{x}}=-\boldsymbol{\widehat{\theta}}\cos\varphi-\boldsymbol{\widehat{\varphi}}\sin\varphi$. Comparing it with Eq.~(\ref{E_plane_wave_components}) yields $\alpha=\varphi+\pi$, although the individual angles $\varphi$ and $\alpha$ are meaningless.

\begin{figure}[!tbh]
\includegraphics[width=9cm]{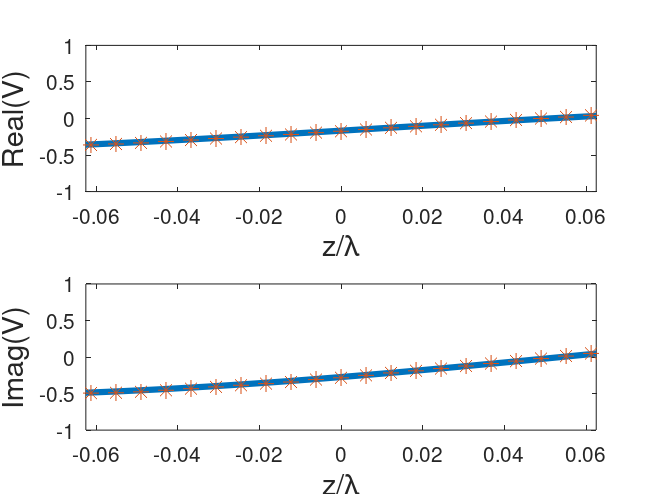}
\caption{Real and imaginary parts of the voltage $V(z)$ for the plane wave incidence shown in Figure~\ref{left_incidence}, at frequency 30MHz or $L/\lambda=1/16$. The continuous line is the analytic solution and the stars are the ANSYS simulation results.}
\label{Ex_th_pi_30M}
\end{figure}
\begin{figure}[!tbh]
\includegraphics[width=9cm]{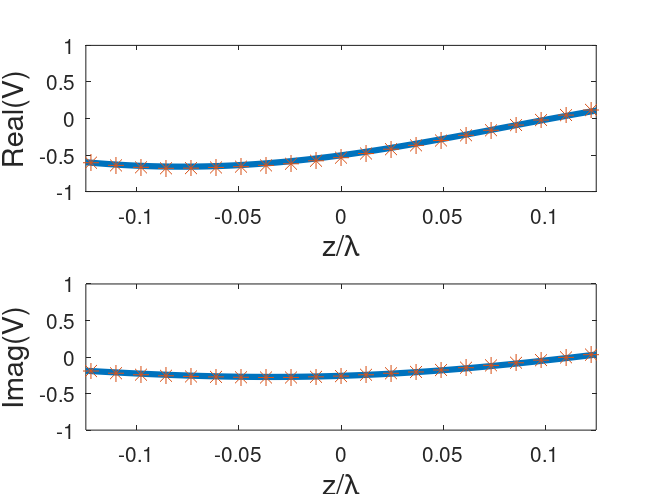}
\caption{Same as Figure~\ref{Ex_th_pi_30M}, for frequency 60MHz or $L/\lambda=1/8$.}
\label{Ex_th_pi_60M}
\end{figure}
\begin{figure}[!tbh]
\includegraphics[width=9cm]{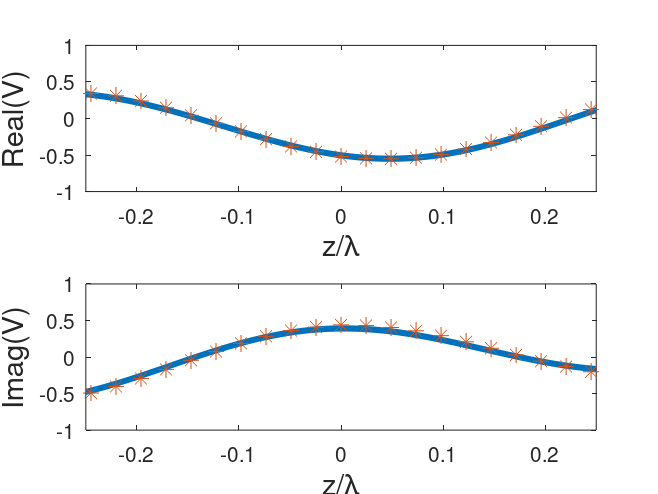}
\caption{Same as Figure~\ref{Ex_th_pi_30M}, for frequency 120MHz or $L/\lambda=1/4$.}
\label{Ex_th_pi_120M}
\end{figure}

Figures~\ref{Ex_th_pi_30M}-\ref{Ex_th_pi_120M} show the voltage for the $\mathbf{\widehat{x}}$ polarised
plane wave from $\theta=\pi$ in Figure~\ref{left_incidence} for frequencies 30, 60 and 120MHz, or $L/\lambda=1/16$,
1/8 and 1/4, respectively.

\begin{figure}[!tbh]
\includegraphics[width=9cm]{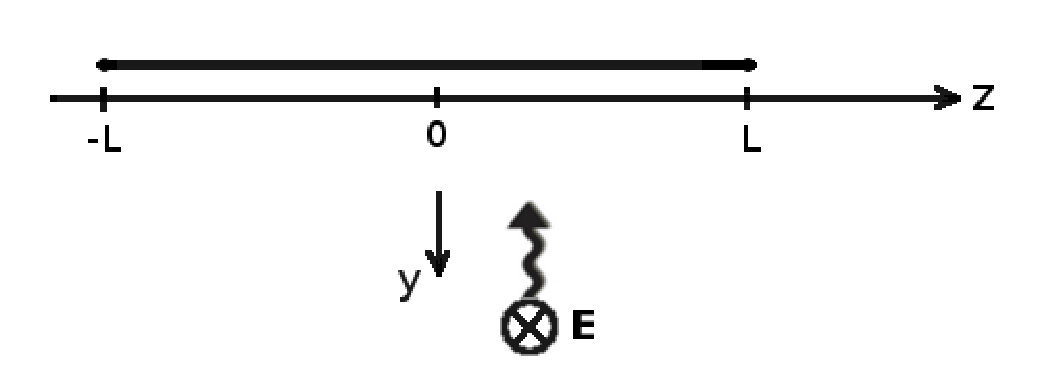}
\caption{Matched TL illuminated by a $-\mathbf{\widehat{x}}$ polarised plane wave from $(\theta=\pi/2,\varphi=\pi/2)$. The view is from the positive $x$ axis direction, so that only the ``upper positive'' conductor is seen.}
\label{front_incidence}
\end{figure}

In the next example we use a plane wave hitting from $\theta=\varphi=\pi/2$ with
phase $e^{jky}$, polarized in the $-\mathbf{\widehat{x}}$ direction, as shown in Figure~\ref{front_incidence}.

\begin{figure}[!tbh]
\includegraphics[width=9cm]{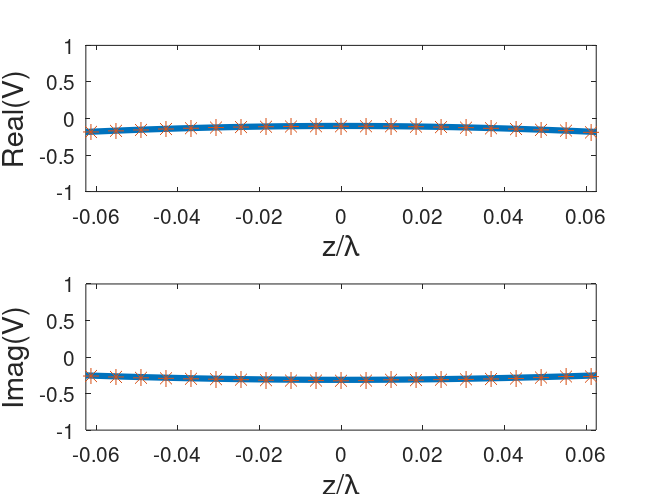}
\caption{Real and imaginary parts of the voltage $V(z)$ for the plane wave incidence shown in Figure~\ref{front_incidence}, at frequency 30MHz or $L/\lambda=1/16$. The continuous line is the analytic solution and the stars are the ANSYS simulation results.}
\label{Ex_th_pi_over_2_phi_over_2_30M}
\end{figure}

\begin{figure}[!tbh]
\includegraphics[width=9cm]{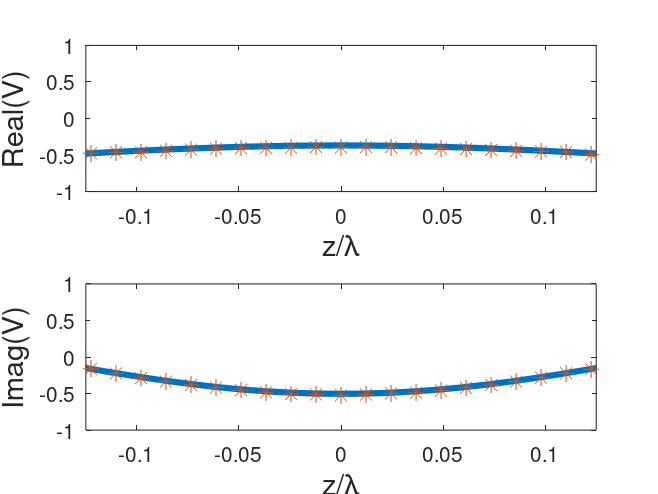}
\caption{Same as Figure~\ref{Ex_th_pi_over_2_phi_over_2_30M}, for frequency 60MHz or $L/\lambda=1/8$.}
\label{Ex_th_pi_over_2_phi_over_2_60M}
\end{figure}

\begin{figure}[!tbh]
\includegraphics[width=9cm]{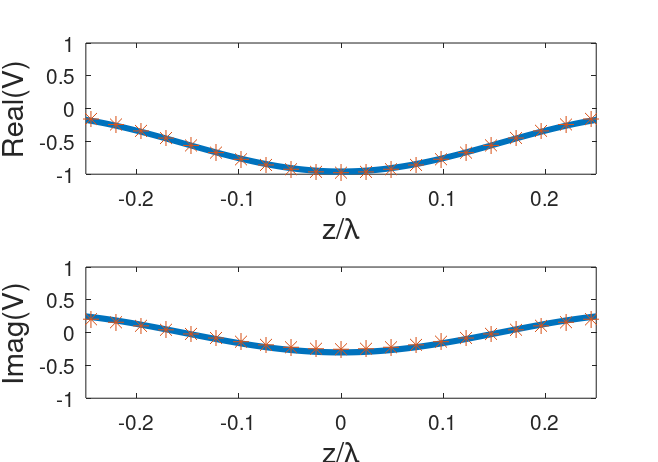}
\caption{Same as Figure~\ref{Ex_th_pi_over_2_phi_over_2_30M}, for frequency 120MHz or $L/\lambda=1/4$.}
\label{Ex_th_pi_over_2_phi_over_2_120M}
\end{figure}

Figures~\ref{Ex_th_pi_over_2_phi_over_2_30M}-\ref{Ex_th_pi_over_2_phi_over_2_120M} show the voltage for the $-\widehat{x}$ polarised plane wave from $\theta=\pi/2$ and $\varphi=\pi/2$ in Figure~\ref{left_incidence} for frequencies 30, 60 and 120MHz, or $L/\lambda=1/16$, 1/8 and 1/4, respectively.

\begin{figure}[!tbh]
\includegraphics[width=9cm]{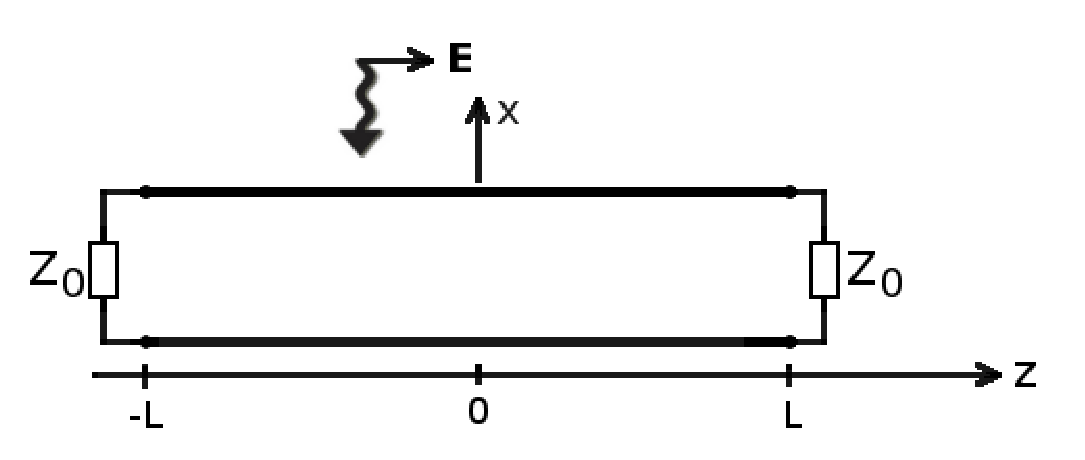}
\caption{Matched TL illuminated by a $\mathbf{\widehat{z}}$ polarised plane wave from $(\theta=\pi/2,\varphi=0)$.}
\label{up_incidence}
\end{figure}

In the next example we use a plane wave incident from $\theta=\pi/2$ and $\varphi=0$, with phase $e^{jkx}$ as shown in Figure~\ref{up_incidence}.

\begin{figure}[!tbh]
\begin{centering}
\includegraphics[width=9cm]{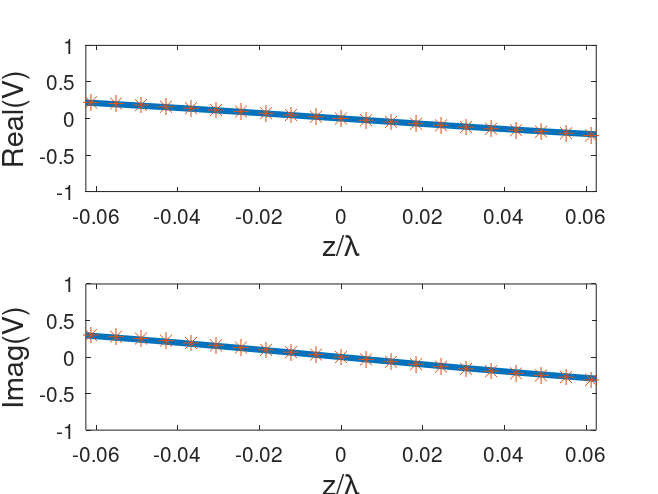}
\caption{Real and imaginary parts of the voltage $V(z)$ for the plane wave incidence shown in Figure~\ref{up_incidence}, at frequency 30MHz or $L/\lambda=1/16$. The continuous line is the analytic solution and the stars are the ANSYS simulation results.}
\label{Ez_th_pi_over_2_phi_0_30M}
\end{centering}
\end{figure}

\begin{figure}[!tbh]
\includegraphics[width=9cm]{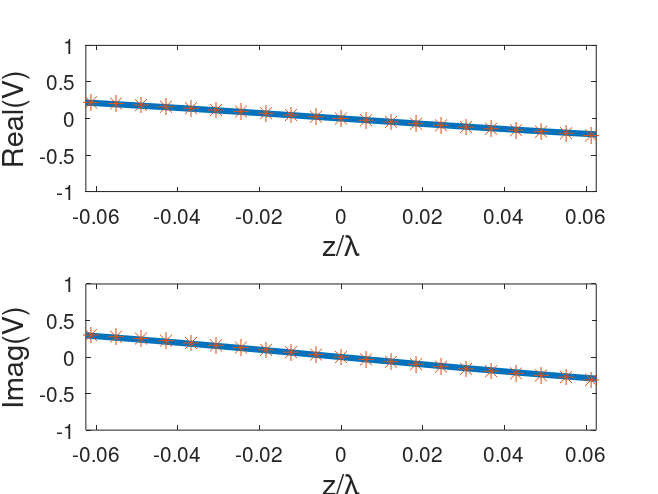}
\caption{Same as Figure~\ref{Ez_th_pi_over_2_phi_0_30M}, for frequency 60MHz or $L/\lambda=1/8$.}
\label{Ez_th_pi_over_2_phi_0_60M}
\end{figure}

\begin{figure}[!tbh]
\begin{centering}
\includegraphics[width=9cm]{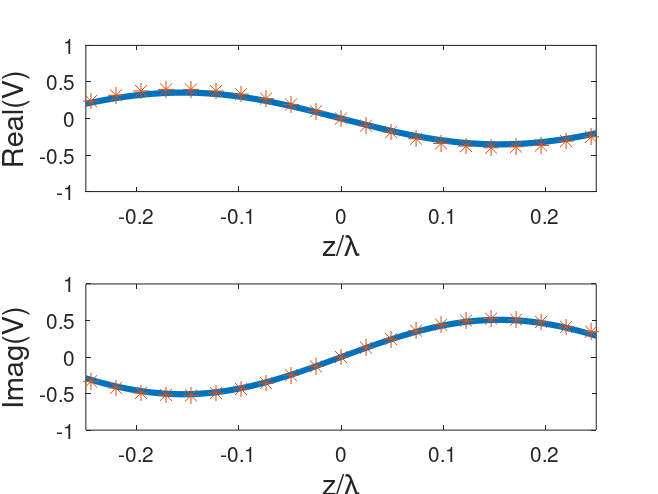}
\caption{Same as Figure~\ref{Ez_th_pi_over_2_phi_0_30M},  for frequency 120MHz or $L/\lambda=1/4$.}
\label{Ez_th_pi_over_2_phi_0_120M}
\end{centering}
\end{figure}

Figures~\ref{Ez_th_pi_over_2_phi_0_30M}-\ref{Ez_th_pi_over_2_phi_0_120M} show the voltage for the $\mathbf{\widehat{z}}$ polarised plane wave from $\theta=\pi/2$ and $\varphi=0$
in Figure~\ref{up_incidence} for frequencies 30, 60 and 120MHz, or $L/\lambda=1/16$,
1/8 and 1/4, respectively.

\subsection{Non matched transmission line}
\label{non_matched}

We compare here several unmatched cases for the $-\widehat{x}$ polarised plane wave from $\theta=\pi/2$, and
$\varphi=\pi/2$, shown in Figure~\ref{front_incidence} for frequency 60~MHz.

\begin{figure}[!tbh]
\includegraphics[width=9cm]{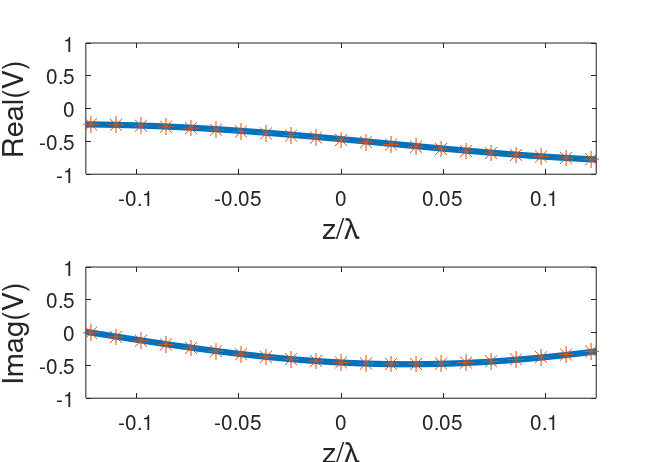}
\caption{Real and imaginary parts of the voltage for a non matched case of $Z_L=Z_0/2$ and $Z_R=2Z_0$ (or $\Gamma_L=-1/3$, $\Gamma_R=1/3$), for the plane wave incidence shown in Figure~\ref{front_incidence}, at frequency 60MHz (or $L/\lambda=1/16$). The continuous line is the analytic solution, and the stars are the ANSYS simulation results.}
\label{Ex_th_pi_over_2_phi_over_2_60M_ZL_half_ZR_2}
\end{figure}

\begin{figure}[!tbh]
\includegraphics[width=9cm]{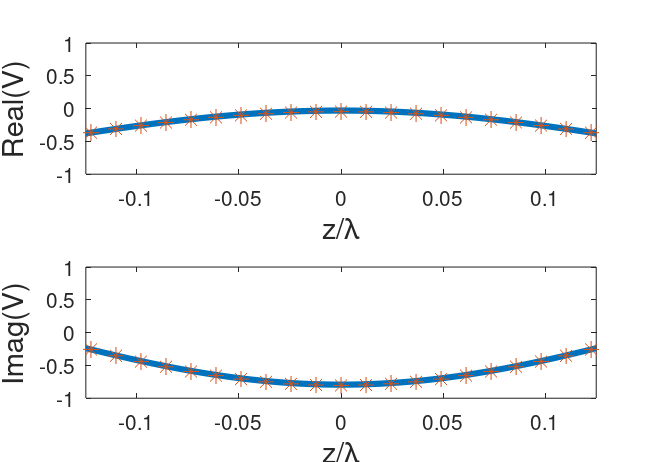}
\caption{Same as Figure~\ref{Ex_th_pi_over_2_phi_over_2_60M_ZL_half_ZR_2} for
$Z_L=Z_R=Z_0/2$ or $\Gamma_L=\Gamma_R=-1/3$.}
\label{Ex_th_pi_over_2_phi_over_2_60M_ZL_half_ZR_half}
\end{figure}

\begin{figure}[!tbh]
\includegraphics[width=9cm]{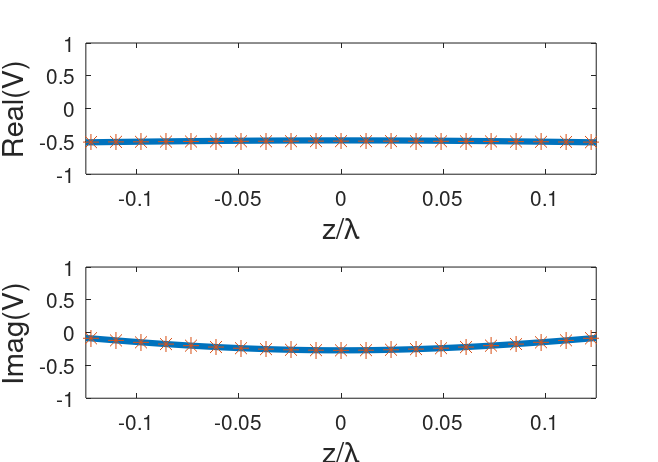}
\caption{Same as Figure~\ref{Ex_th_pi_over_2_phi_over_2_60M_ZL_half_ZR_2} for
$Z_L=Z_R=2Z_0$ or $\Gamma_L=\Gamma_R=1/3$.}
\label{Ex_th_pi_over_2_phi_over_2_60M_ZL_2_ZR_2}
\end{figure}

Figures~\ref{Ex_th_pi_over_2_phi_over_2_60M_ZL_half_ZR_2}-\ref{Ex_th_pi_over_2_phi_over_2_60M_ZL_2_ZR_2}
show the voltage for the cases: $Z_L=Z_0/2$ and $Z_R=2Z_0$, $Z_L=Z_R=Z_0/2$
and $Z_L=Z_R=2Z_0$, respectively.

\section{Conclusion}

In this work we derived analytically the voltage along a two-conductors dielectrically isolated quasi-TEM transmission line of any small electric cross section, connected to passive loads and hit by a monochromatic plane wave, as shown in Figure~\ref{config}.

For this derivation we used our knowledge on the radiation properties of such TL \cite{{QuasiTEM}} to build a generalized S matrix which describes the radiating system and used the reciprocity to derive the voltage induced on the TL.


We validated our analytic results in Section~\ref{validation_hfss} for both the double-matched TL case and the non matched case for different plane wave incidences.


To be mentioned that the formalism elaborated in this work is applicable to any configuration for which one aims to obtain the receiving characteristics from the radiation characteristics, and we already applied it succesfully for the free space TEM transmission lines \cite{receive_TL}, deriving from it the response of free space TL to a monochromatic incident plane wave \cite{receive_TL}.

The algorithm can also be applied to multiconductor transmission lines (MTL). Practical MTL (for which the field lines extend beyond the insulator) of $N$ conductors develop $N-1$ modes. Each mode has its own characteristic impedance and its own propagation delay. In case of uniform medium (air or “infinite” insulator), there are still $N-1$ degenerate modes, all with the same propagation delay, like in \cite{Paul}. We dealed in the past with MTL \cite{MTL, lossy_MTL}, developing a robust algorithm to obtain the properties of the modes. It is reasonable to treat radiation/absorption of MTL using a per mode analysis for the reciprocity between radiation and absorption.


%



\section*{Acknowledgment}
This work has been partially supported by the Israeli Science Foundation (ISF)

\ifCLASSOPTIONcaptionsoff
  \newpage
\fi

\newpage





\end{document}